\documentclass{PoS}

\title{Upsilon and $\chi_b$ analyses at CLEO}

\ShortTitle{Upsilon and $\chi_b$ analyses at CLEO}

\author{\speaker{Jean Duboscq} for the CLEO Collaboration
        \thanks{ I am thankful to the National Science Foundation for its
 essential support of this work, as well as to the staff of CESR for its
 excellent accelerator work.}\\ 
        Cornell University, USA\\ 
        E-mail: \email{jed@mail.lepp.cornell.edu}}

\abstract{I detail recent work done by the CLEO collaboration using 
data gathered at the CESR accelerator concerning the $\Upsilon$ system. 
Results include $B_S$ production at the $\Upsilon(5S)$, decays of the 
$\Upsilon(nS)$ (n=1,2,3) to $\tau$ pairs, the $\Upsilon$ di-electron 
width, decays of the $\Upsilon$ to two hadrons and a photon, and the 
observation of the first di-pion transition between $\chi_B$ states}

\FullConference{International Europhysics Conference on High Energy Physics\\
		 July 21st - 27th 2005\\
		 Lisboa, Portugal}

\begin{document}


\section{The $\Upsilon$ System}
The $\Upsilon(nS)$ system, a bound state of a $b$ and $\overline{b}$ 
quark, is produced in the CLEO detector in collisions of electrons and 
positrons from the CESR accelerator. When $n \ge 4$ the state decays 
to $B$ mesons. For $n<4$, below $B$ meson threshold, the two quarks 
must annihilate to produce non-B final states. These annihilations are 
mediated either by a virtual photon, the exchange of two gluons, or 
the radiation of one photon and emission of a photon. These states can 
also transition to other $n^{2s+1}L_J$ states via emission of, eg, a 
photon or two pions. These other states are the QCD analog of positronium. 
Since the $b$ quark is quite heavy, non-relativistic quantum mechanics 
can be used to understand the spectroscopy and dynamics. It is interesting 
to note that the catalogue of final states is incompletely known - for 
instance in the case of the $\Upsilon(1S)$ less than $10 \%$ of all 
final states are measured~\cite{PDG}. The $\Upsilon$ system also supplies 
a laboratory to study $b$ quarks in a setting different from $B$ meson 
decays. For instance, one's confidence in Lattice QCD (LQCD) as applied 
to the extraction of CKM matrix elements in B meson decays, would be 
boosted if LQCD's prediction of the $\Upsilon$ system were in line with 
experimental results.

\section{ The CLEO III  Detector and Data Sample}
The CLEO III detector\cite{CLEODET} provides particle measurements over a large solid 
angle. This detector, in
 a 1.5 T magnetic field, includes a silicon 
tracker, and a drift chamber for charged particle detection, as well 
as $dE/dx$ particle identification. Outside this detector is a Ring 
Imaging Cerenkov (RICH) detector, providing excellent separation of 
$\pi$, $K$, $p$. Muon chambers round out the subsystems, providing $\mu$ 
detection above momenta around 1.5 GeV.
 This detector sits on the CESR 
accelerator, which provides electron positron collisions near $E_{CM} 
\approx 10 GeV$, with symmetric beam energies. 
 CLEO III has the world's 
largest data sample for the $\Upsilon$ resonances below $B$ meson threshold. 
There are approximately 20 million $\Upsilon(1S)$, 10 million $\Upsilon(2S)$ 
and 5 million $\Upsilon(3S)$ decays in the data set. CLEO III has also 
collected $0.42 fb^{-1}$ of data in the $\Upsilon(5S)$ region.
 
 \section{ $B_S$ Production at the $\Upsilon(5S)$}
  The $\Upsilon(5S)$ is above 
the threshold for producing $B_S$ mesons, and should decay, among other 
things, to $B_S^{(*)} \overline{B}_S^{(*)}$ states. This would be an 
interesting place for B factories to run to investigate $B_S$ mixing. 
Up to now, no experiment has reported the
   observation of $B_S$ production 
in $\Upsilon$ decays - observation and measurement of this
    would 
be an invaluable input to planning for $B_S$ production at the B factories.
 
  An elementary estimation leads to the hypothesis that the $B_S$ meson 
decays to final states with a $D_S$ with a branching fraction of  $92 
\pm 11\%$. Non strange $B$ meson decays contain far fewer $D_S$ mesons. 
Thus CLEO maps out the production of $D_S$ mesons in the $\phi \pi$ 
final states as a function of $D_S$ meson beam energy scaled momentum 
using off resonance and on resonance $\Upsilon(4S)$ data, and on resonance 
$\Upsilon(5S)$ data.
   The $\Upsilon(4S)$ off and on resonance yields 
are then scaled to the $\Upsilon(5S)$ energies, to account for continuum 
and non strange $B$ meson production of the $D_S$, which is then subtracted 
from the observed spectrum. The remainder is attributed to the decay 
of $B_S$ mesons to the $D_S$, allowing us to observe for the first time 
the production of the $B_S$ at the $\Upsilon(5S)$, and quote:
    $ 
Br( \Upsilon(5S) \to B_S^{(*)} \overline{B}_S^{(*)}) = 16.0 \pm 2.6 
\pm 6.3 \% $. This work is now published in \cite{BSPROD}.
    
    A second search attempts to identify exclusive $B_S$ final state. 
CLEOIII searched for the $B_S$ final states $\psi \phi$, $\phi \eta$, 
$\psi \eta'$ and $D_S^{(*)} \pi^-$, $D_S^{(*)} \rho^-$.
    In a plane defined by the difference in energy between the beam 
energy and the final state particle energy versus the beam constrained 
mass of the final states, kinematics dictate a separation of the $B_S 
\overline{B}_S$, $B_S \overline{B}_S^*$, and $B_S^* \overline{B}_S^*$ 
states. CLEO III observes a preponderance of events in the $\psi X$ 
and $D_S^{(*)} X $ modes in the $B_S^* \overline{B}_S^*$ region. This indicates 
that the $B_S^* \overline{B}_S^*$ dominates the decay of the $\Upsilon(5S)$ 
as expected in the Unitarized Quark Model \cite{Ono}. Results from 
this work are detailed in \cite{BStar}.
    
    \section{ $\Upsilon$ Decays to Two Leptons}
    
    The decay of the $\Upsilon(nS)$ (n=1,2,3) to two leptons is interesting 
because  it is a probe of the coupling of two $b$ quarks to a virtual 
photon. As such it can be used as a test bed for LQCD \cite{Davies}. 
In addition, this simple decay is a good place to test lepton universality 
by comparing the final states $ee$, $\mu\mu$, $\tau\tau$. Phase space 
corrections should be very small, and thus branchings fractions to these 
states should be equal. A deviation from this expectation could be a 
manifestation of new physics \cite{Sanchis}. 
    The decays of the $\Upsilon(1S)$ to lepton pairs are currently measured 
at the $2$ to $ 5\%$ level. The $\Upsilon (2S) $ decays to the $ee$ 
and $\mu\mu$ final states are also measured to approximately $8\%$, 
while the $\tau\tau$ final state is known only to within $100\%$. For 
the $\Upsilon(3S)$, only the $\mu\mu$ final state is well know to an 
error of $10\%$, while the $ee$ mode is known to exist (through the 
production of the 3S at colliders) and the $\tau\tau$ mode has not been 
observed.
    
    \section{Analysis of $\Upsilon \to \tau\tau$}
     The analysis of the decay $\Upsilon \to \tau \tau$ follows the 
method used in the analysis of the $\Upsilon \to \mu\mu$ decay\cite{Istvan}.
 The data on and off resonance for the n=1,2,3 $\Upsilon(nS)$ 
states are skimmed for two track events and selection criteria are tuned 
 using data from the $\Upsilon(4S)$ to isolate both the $\mu\mu$ and 
$\tau \tau$ final states ($\tau$ leptons decay to final states containing 
one charged track approximately $75\%$ of the time.) The off resonance 
data are scaled according to luminosity and beam energy (to account 
for background evolution) and subtracted from the on resonance data. 
The excess is attributed to $\Upsilon$ decays. The ratio of the $\tau\tau$ 
and $\mu\mu$ sample, in which interference between on and off resonance 
production cancels, provides, after efficiency correction, for a direct 
test of lepton universality. In addition, this method allows as a cross 
check the verification that the decay of the $\Upsilon(4S)$ to these 
final states does not occur (at our level of sensitivity.)  Clear excesses 
of events are observed in both $\mu\mu$ and $\tau\tau$ modes at the 
$\Upsilon(nS)$, n=1,2,3, and no excess is observed at the $\Upsilon(4S)$. 
    The observed event yields in both final states are corrected for 
efficiency, and, for the 2S and 3S, cascade decays to lower $\Upsilon$ 
states, with reasonable assumptions for unmeasured decays. The interference 
corrected number of $\mu\mu$ events is in perfect agreement with our 
previous measurement of $Br(\Upsilon \to \mu\mu)$, and in addition, 
the off resonance production of $\mu\mu$ and $\tau\tau$ final states 
is in complete agreement with the Standard Model expectation.
    The result of the analysis is that $R= {Br(\Upsilon\to\tau\tau )
}/ {Br(\Upsilon \to \mu\mu)}$ is measured as follows:
    \begin{eqnarray}
      R(1S) & = & 1.06 \pm 0.02 \pm 0.00 \pm 0.03 \\ 
         R(2S) & = & 1.00 \pm 0.03 \pm 0.12 \pm 0.03 \\
           R(3S)  & = &1.05 \pm 0.07 \pm 0.05 \pm 0.03 
           \end{eqnarray}
           where the quoted errors are due to statistics, cascade modeling 
and other systematics, respectively. The systematic errors for this 
preliminary result are quite conservative and should decrease in the 
course of the analysis. 
     This is a first observation of the decay 
$\Upsilon(3S) \to \tau\tau$ at the 10 $\sigma$ statistical level, and 
is a marked improvement in the error on the branching fraction for $\Upsilon(2s) 
\to \tau\tau$. 
     Note that although this result does seem to agree with the expectations 
from lepton universality, the deviation in a model such as that proposed 
by \cite{Sanchis} in which the $\Upsilon$ might occasionally radiatively 
decay to an $\eta_b$, which then decays via a (non-standard) Higgs, will 
 be sensitive to the energy of the photon. The efficiency for seeing the 
 final $\tau\tau$ 
state will be lower in this scenario because of the extra photon, and 
will dilute the power of the result reported here, depending on the 
photon energy.
     
     \section{ Dielectron Width of the $\Upsilon$ }
     The dielectron width of the $\Upsilon$ probes the coupling of the 
$b\overline{b}$ system to the  electron-positron system. Ratios of these 
width among the different $\Upsilon$ states are of interest to those 
who study LQCD as a probe of the quark anti-quark wave function overlap 
as these should be calculable. The direct measurement of the production 
of $e^+e^-$ in $\Upsilon$ decay is hindered by the large background 
due to Bhabha scattering that must be substracted in a direct analysis. 
At CLEO, the time inverted process can be measured by measuring the 
line shape for the production of $\Upsilon$ as a function of the center 
of mass energy of electron positron collisions. The total area of this 
line shape is the desired width. The observed line shape is a convolution 
of several components, including the inherent width $\Gamma_{ee}$ ($\approx 
1 keV$) , the accelerator beam energy spread ($\approx 4 MeV$), and 
the effects of initial state radiation, which also provides a non Gaussian 
smearing on the order of MeVs. The initial state radiation function 
smears the observed line shape assymetrically, giving a large tail on 
the high end of $s^{1/2} = E_{CM}$. 
     The method used in this scan 
over the $\Upsilon(nS)$, n=1,2,3 involved weekly scans over the resonances, 
including off resonance points 20 MeV below the resonance peak to gauge 
continuum backgrounds. In order to probe the stability of the beam energy 
measurement, in each scan, the point with the largest slope with respect 
to $s^{1/2}$ was revisited. The selected events were hadronic, as these 
provided large well understood triggering and reconstruction efficiencies. 
The line shape was fit to a $1/s$ shape to account for continuum hadron 
production, as well as the convoluted $\Upsilon$ line shape. Included 
in this fit was an estimation of the contribution due to the interference 
of the continuum hadron production with $\Upsilon$ production. The number 
of events observed was corrected for efficiency, $\tau$ pair feedthrough, 
as well as for the far smaller contribution due to cosmic rays and beam 
gas collisions. The efficiency correction included a correction for 
the effective invisible width due to both trigger inefficiency and real 
physics, as estimated using the cascade $\Upsilon(2S) \to \Upsilon(1S) 
\pi^+\pi^-$, and Monte Carlo simulations. 
     The resulting preliminary fits measure the widths to be:
     \begin{eqnarray}
     \Gamma_{ee}(\Upsilon(1S)) & = & 1.336 \pm 0.009 \pm 0.019 \, keV  \\
      \Gamma_{ee}(\Upsilon(2S)) & = & 0.616 \pm 0.010 \pm 0.009 \, keV \\
\Gamma_{ee}(\Upsilon(3S)) & = & 0.425 \pm 0.009 \pm 0.006 \, keV  
\end{eqnarray}
 The errors in the above are statistical and systematic. The systematic 
error includes a $1.3\% $ error due to the knowledge of the luminosity. 
The ratios that are of direct relevance to LQCD are:
 \begin{eqnarray}
  {  \Gamma_{ee}(\Upsilon(2S))  \over{ \Gamma_{ee}(\Upsilon(1S))} } & = & 0.461 \pm 0.008 \pm 0.003 \\
 {  \Gamma_{ee}(\Upsilon(3S))  \over{ \Gamma_{ee}(\Upsilon(1S))}} & =  & 0.318 \pm 0.007 \pm 0.002 \\
   {  \Gamma_{ee}(\Upsilon(3S))  \over{ \Gamma_{ee}(\Upsilon(2S))} } & = & 0.690 \pm 0.019 \pm 0.006
   \end{eqnarray}
   
   \section{Direct Photons in $\Upsilon$ Decays}
  Below $B$ meson threshold, two important channels in $\Upsilon$ decay 
are the three gluon $ggg$  and the $\gamma g g$ exchange diagrams. The 
photon in the $\gamma gg$ process escapes into the detector, and thus 
is a direct witness of the physics of the $b\overline{b}$ overlap. The 
ratio of the number of events observed in these two processes
   $R_\gamma 
= N(\gamma gg ) / N(ggg) $ is a function of the strong and electromagnetic 
coupling, as well as the charge of the b quark, and can be used to gauge 
the relative importance of these two processes. 
    The search proceeds 
by tabulating hadronic events with an isolated high energy photon reconstructed 
in CLEO. The energy spectrum of the photon is used to untangle the physics 
from the backgrounds.  Continuum processes are subtracted by an off 
resonance subtraction. The two main backgrounds are photons that come 
from ISR, which are the dominant background at the highest photon energy, 
while at lower energies the dominant background comes from photons resulting 
from $\pi^0$ decays. The ISR background is expected to be reasonably 
well simulated by the event simulation. The background due to $\pi^0$'s 
comes from complicated physics processes, and one might question the 
veracity of a Monte Carlo model. To circumvent this, CLEO uses a data 
driven Monte Carlo : using the reconstructed spectrum of charged pions, 
we can predict the spectrum of real $\pi^0$, and then throw Monte Carlo 
$\pi^0$'s that reproduce this spectrum. These $\pi^0$'s are then allowed 
to simulate the real background process. The ISR and $\pi^0$ derived 
background are subtracted from the continuum subtracted on resonance 
data, and the resulting distribution is fit to a sum of two contributions. 
The first represents the mismatch of our subtraction at low photon energies, 
while the second component is the distribution expected for the direct 
photons. We fit to two models \cite{Field}\cite{Garcia}, both of which 
expect a predominance of photons at higher energies. These models are 
derived for the $\Upsilon(1S)$ but lacking other suitable models, we 
also apply them to the $2S$ and $3S$ states. The fits for the higher 
resonances include the effects of cascades to lower resonances.
    In the ratio for $R_\gamma$, the denominator is determined from the 
known number of $\Upsilon$ in the CLEO dataset, as well input from the 
PDG\cite{PDG} and Monte Carlo simulations. 
     The preliminary results are:
     $ R_\gamma(1S) = 2.90 \pm 0.007 \pm 0.22 \pm 0.15 \% $, 
     $   R_\gamma(2S) = 3.49 \pm 0.03 \pm 0.58 \pm 0.18 \%$, and
     $    R_\gamma(3S) = 2.88 \pm 0.03 \pm 0.38 \pm 0.12 \% $
          where the errors are statistical, systematic and model dependence. 
The systematic error is dominated by the $\pi^0$ photon feedthrough 
model. 
          The value of $R_\gamma (1S)$ is consistent with previous values, 
and has similar systematic errors, but smaller statistical errors. This 
is the first measurement of $R_\gamma(2S)$ and $R_\gamma(3S)$.

          \section{ The Decay $\Upsilon(1S) \to h^+ h^- \gamma $}
          The radiative decays of the $\Upsilon(1S)$ can be used to 
probe the 2 gluon structure. There are many interesting results from 
$J/\psi$ decays, including the observation of tensor states $f_2(1270)$ 
\cite{PLUTO, Scharre, CrystalBall, DMTwo}
and $f_2'(1525)$ \cite{MARKIII, BEStensor},
 the claim of a glueball candidate known as the $f_j(2220)$ 
\cite{BESFJ}, 
and the observation of a $p\overline{p}$ near threshold enhancement 
termed the $X(1860)$ \cite{BES}.  In the $\Upsilon(1S)$ system, one might 
hope to observe similar effects suppressed by a factor due to the relative 
quark charge and the quark mass for the propagator, squared, which roughly 
predicts rates reduced by a factor of $1/40$. 
           In this analysis, 
CLEO searches for the final state $h^+h^-\gamma$ in $\Upsilon(1S)$ decays, 
by requiring a bachelor photon of energy greater than 4 GeV. The $h^{\pm}$ 
are identified as $\pi$, $K$ or $p$ by combining dEdx and RICH information. 
In addition, the knowledge of the total beam energy is used as an extra 
input to test for PID hypothesis consistency. Continuum decays are accounted 
for by subtracting suitably scaled off resonance data. 
            The resulting 2 body mass spectra reveal a clear $f_2(1270)$ in the 
$\pi\pi$ spectrum, as well as a clear $f_2'(1525)$ in the $KK$ spectrum. 
No clear structure is observed in the $p\overline{p}$ spectrum. No evidence 
for the $f_j(2220)$ is observed in any of the spectra. There is an unexplained 
broad excess of events in the $KK$ mass distribution above 2 GeV. In 
addition, there is no evidence of the $X(1860)$ in the $p\overline{p}$ 
spectrum. The decay angles are fit and it is confirmed that the $f_2(1270)$ 
and the $f_2'(1525)$ are spin 2 objects decaying predominantly with 
helicity 0.
         Preliminary branching ratios are:
         $ Br(\Upsilon(1S) \to \gamma f_2(1270)) = 10.2 \pm 0.8 \pm 0.7 \times 10^{-5}$ and 
         $ Br(\Upsilon(1S) \to \gamma f_2'(1525)) = 3.7 ^{+0.9}_{-0.7} \pm 0.8 \times 10^{-5}$.
           The excess in the $KK$ spectrum for $2 GeV < m_{KK} < 3 GeV$ is $Br(\Upsilon(1S) \to \gamma K^+K^-) = 1.14 \pm 0.08 \pm 0.10 \times 10^{-5}$. 
            Upper limits at $90 \%$ C.L. are set at:
            $Br(\Upsilon \to \gamma f_2(980)) < 3 \times 10^{-5} $, 
            $ Br(\Upsilon \to \gamma f_2(2050)) < 0.6 \times 10^{-5} $,
$Br(\Upsilon \to \gamma f_0(1710)) < 0.7 \times 10^{-5} $. 
For the mass region $2 GeV < m_{p\overline{p}}< 3 GeV$, $Br(\Upsilon(1S) 
\to \gamma p \overline{p}) < 0.6 \times 10^{-5}$. 
Additional limits on the branching fraction to the $f_j(2220)$ and the 
$X(1860)$ are set at the $10^{-6}$ level.
The branching fractions for 
the    $ f_2(1270)$ and $f_2(1270)$ are consistent with the naive expectation 
derived from $J/\psi$ decays. It is not clear that any of the other 
results are in conflict with this scaling. The results of this work can be found in \cite{hhgamma}. An additional analysis is 
underway looking for final states including a photon and $\pi^0\pi^0$, 
$\eta\eta$, and $\pi^0\eta$. 

\section{Observation of $\chi_b(2P) \to \chi_b(1P) \pi^+\pi^-$}
 The transition of $\chi_b(2P) \to \chi_b(1P) \pi^+\pi^-$ is observed 
in an electromagnetic decay of the $\Upsilon(3S)$ to the $\chi_b(2P)$ 
followed by the dipion transition down to the $\chi_b(1P)$ followed 
by a electromagnetic transition down to the $\Upsilon(1S)$. The $\Upsilon(1S)$ 
is then observed by its decay to $ee$ or $\mu\mu$. The final state, 
excluding the leptons, consists of 
  2 charged pions and 2 photons, which is also the final state of the 
chain $\Upsilon(3S) \to \Upsilon(2S) \pi^+\pi^-$, $\Upsilon(2S)\to \chi_b 
\gamma$, $\chi_b \to \Upsilon(1S) \gamma$. This measured background 
with slightly different photon energies and dipion mass can be used 
to gauge the correctness of the analysis of the sought after transition. 
Other backgrounds exist but require the loss of at least 1 photon, and 
are thus somewhat suppressed. The analysis plots the energy of largest 
energy photon against the inferred mass of the dipion system.
   Two 
analyses were pursued. The cascade pions tend to have low momentum, 
and thus, reconstruction of both of them tends to have low efficiency. 
The Di-pion analysis attempts to 
    reconstruct both pions. In this analysis, the $\Upsilon(2S)$ contamination 
is found to be consistent with expectation, and 7 events are observed 
in the signal region over a background of 1.2 expected events. The efficiency 
for this analysis is $\epsilon= 4.5\%$. 
     The single pion analysis 
sees 17 events, over an expected background of 3.3 events, with an efficiency 
of $\epsilon = 8.5\%$. 
      The two analyses see consistent results and can be combined to 
give a 6 $\sigma$ significant observation. CLEO thus reports the first 
observation of a dipion transition outside of a $^3S_1$ system, and 
a preliminary width of 
      $ \Gamma( \chi_b(2P) \to \chi_b(1P) \pi^+\pi^-) \approx 0.9 keV $.
 Further details can be found in \cite{Chib}.
    
\section{Conclusions}
This talk has highlighted results from the $\Upsilon$ analyses at CLEOIII. 
CLEO has made the first observation of $B_S$ production in $\Upsilon(5S)$ 
decays. CLEO has also for the first time observed the decay $\Upsilon(3S)\to 
\tau\tau$, and obtained precision results on the ratio of the
 $\tau\tau$ 
width to the $\mu\mu$ width in $\Upsilon(nS)$ decay, n=1,2,3. A  measurement 
of the dielectron width of the $\Upsilon(nS)$ n=1,2,3 was also presented. 
Direct photon measurements in hadronic $\Upsilon$ decays presented here 
shed light on the relative importance of $\gamma gg$ and $ggg$ intermediate 
states in $\Upsilon$ decays. The substructure of the hadrons in the 
 $\Upsilon(1S) \to \gamma h^+ h^-$ decays was determined to be consistent 
with the extrapolated expectation from $J\psi$ decays. A broad enhancement 
in the $KK$ channel was observed, and no evidence was seen for the $f_j(2220)$ 
or the $X(1600)$. 
  The first observation of the dipion transition between the $\chi_b(2P)$ 
and the $\chi_b(1P)$ was also presented. 
  All results presented in this talk are preliminary unless otherwise 
noted.

\end{document}